\documentclass{aa}
\usepackage{graphicx}
\usepackage{natbib}
\usepackage{longtable}
\begin{document} 
\title{A photometric long-term study of CP stars in open clusters}
 
\author{E.~Paunzen\inst{1} 
\and H.~Hensberge\inst{2}
\and H.M.~Maitzen\inst{1}
\and M.~Netopil\inst{1,3}
\and C.~Trigilio\inst{4} 
\and L.~Fossati\inst{1,5}
\and U.~Heiter\inst{6}
\and M.~Pranka\inst{1}}


\institute{Institut f{\"u}r Astronomie der Universit{\"a}t Wien, T{\"u}rkenschanzstr. 17, A-1180 Wien, Austria
\email{Ernst.Paunzen@univie.ac.at}
\and Royal Observatory of Belgium, Ringlaan 3, Brussels, Belgium
\and Hvar Observatory, Faculty of Geodesy, University of Zagreb, Ka\v{c}i\'{c}eva 26, HR-10000 Zagreb, Croatia
\and INAF - Osservatorio Astrofisico di Catania, Via S. Sofia 78, 95123 Catania, Italy
\and Department of Physics and Astronomy, The Open University, Milton Keynes, MK7 6AA, United Kingdom
\and Department of Physics and Astronomy, Uppsala University, Box 516, SE-75120 Uppsala, Sweden}

\date{}

\abstract
{Photometric variability of chemically peculiar (CP) stars of the upper main 
sequence is closely connected to their local stellar magnetic field and their
rotational period. Long term investigations, as presented here, help us to identify
possible stellar cycles (as in the Sun). Furthermore, these data provide a basis for
detailed surface mapping techniques.}
{Photoelectric Str{\"o}mgren $uvby$ time series for 27 CP
stars within the boundaries of open clusters are presented. In addition, 
Hipparcos photometric data (from 1989 to 1993) are used for our analysis. Our observations cover a time period of
about six years (1986 to 1992) with typically fifteen measurements for each objects. These
observations help us to determine the rotational periods of these objects.}
{A standard reduction procedure was applied to the data. When possible,
we merged our data sets with already published ones to obtain a more significant result.
A detailed time series analysis was 
performed, involving five different methods to minimize spurious detections.}
{We established, for the first time, variability for fourteen CP stars. For additional two stars,
a merging of already published data sets, resulted in more precise periods, whereas
for six objects, the published periods could be confirmed.
Last, but not least, no significant variations were found for five stars. Apart from 
six stars, all targets seem to be members of their host open clusters.}
{The present observations fill an important gap in previous photometric long-time studies of
CP stars. The presented open cluster members are excellent targets for follow-up 
observations, employing for example polarimetric, high-resolution spectroscopic, and surface mapping 
techniques.}
\maketitle

\begin{table*}
\begin{center}
\caption[]{Results for the observed sample of stars in eleven open clusters. 
}
\begin{tabular}{llrrrccrrcc}
\hline
\hline
Cluster & Name & W & \multicolumn{1}{c}{$V$}   & \multicolumn{1}{c}{$\Delta$a}   & 
Spec. & \multicolumn{1}{c}{Mem} & \multicolumn{1}{c}{$\Delta t$} & Num & \multicolumn{1}{c}{Period} 
& \multicolumn{1}{c}{Period$_{\rm Lit}$}\\
& & & \multicolumn{1}{c}{[mag]} & \multicolumn{1}{c}{[mmag]} &     
& & \multicolumn{1}{c}{[d]} & & \multicolumn{1}{c}{[d]} & \multicolumn{1}{c}{[d]} \\
\hline
Cr 132 & HD 56343 \tablefootmark{a}& 27 & 9.25 & +35 & CP & $-$ & 675.2737 & 15 & 3.25 \\
       & HD 56046 & 10 & 7.59 & & B6\,V \\
\smallskip
       & HD 56162 & 12 & 7.80 & & A1\,V \\
IC 4725 & BD$-$19\,5044L & 98 & 10.20 & +16 & CP & p & 7.0674 & 5 & variable \\
        & HD 170836 & 167 & 8.95 & +46 & CP & y & 1819.1000 & 10 & 0.9(?) \\
        & HD 170860A & 153 & 9.40 & +22 & CP & p & 1819.1000 & 10 & variable \\
        & HD 170719 & 91 & 8.10 & & B5\,III \\
        & HD 170820 & 150 & 7.38 & & K0\,III \\
\smallskip
        & HD 170682 & 50 & 7.95 & & B5\,II/III \\
NGC 2516 & CP$-$60\,944 & 112 & 9.85 & +21 & CP & y & 337.1385 & 17 & 1.10 \\
         & HD 65712 & 230 & 9.86 & +70 & CP & p & 9.1297 & 13 & 1.88 & 1.943\tablefootmark{1} \\
         & HD 66295 & 26 & 9.10 & +45 & CP & y & 9.1303 & 11 & 2.45 & 2.45004\tablefootmark{2} \\
         & CP$-$60\,947 & 13 & 8.06 & & B8\,V \\
\smallskip
         & CP$-$60\,952 & 9 & 9.30 & & B9\,V \\
NGC 3114 & HD 87240 & 25 & 8.65 & +30 & CP & p & 1734.3087 & 28 & variable \\
         & HD 87405 \tablefootmark{a}& 108 & 8.50 & +17 & CP & ? & 1734.2957 & 28 & 2.113 \\
         & HD 87752 & 211 & 9.82 & +6 & CP & y & 4.9909 & 5 & constant \\
         & HD 304841 & 80 & 9.99 & +36 & CP & p & 349.1259 & 21 & 0.6978 \\
         &           &    &      &     &    & &          &    & longer\tablefootmark{b} \\
         & HD 304842 & 234 & 9.72 & +19 & CP & y & 1394.2079 & 15 & 2.31808 & 2.31873\tablefootmark{2} \\
         & HD 87137 & 8 & 8.30 & & B8\,III \\
\smallskip
         & HD 87349 & 86 & 8.62 & & B8\,V \\
NGC 3228 & HD 89856\tablefootmark{a} & 3 & 9.09 & +32 & CP & n & 1734.2579 & 28 & 4.556 & 4.565\tablefootmark{3} \\
         & HD 298053 & 16 & 10.69 & +15 & CP & p & 97.8085 & 21 & 0.84 or 0.94 \\
         & HD 89900 & 5 & 8.19 & & A0\,IV \\
\smallskip
         & HD 89915 & 6 & 7.90 & & A0\,V \\
NGC 3532 & HD 96040 & 413 & 10.00 & +60 & CP & p & 1392.1704 & 17 & 1.13 \\
         &          &     &       &     &    &      &           &    & longer\tablefootmark{b} \\
         & HD 96729 & 449 & 10.00 & +65 & CP & n & 11.9983 & 12 & 1.39 \\
         & HD 96011 & 467 & 9.05 & & A0\,V \\
\smallskip
         & HD 96226 & 409 & 8.03 & & B8 \\
NGC 5460 & HD 122983 & 142 & 9.90 & +20 & CP & p & 1392.1247 & 17 & 3.4\,$<$\,P\,$<$3.8 \\
         & HD 123225 & 55 & 8.85 & +14  & CP & y & 1385.1544 & 10 & constant \\
         & HD 123202 & 79 & 8.90 & & B9\,IV/V \\
\smallskip
         & HD 123224 & 50 & 8.00 & & B8\,II \\
NGC 5662 & CP$-$56\,6330 & 85 & 10.89 & +33 & CP & n & 92.7977 & 17 & 0.81 \\
         & HD 127924\tablefootmark{a} & 187 & 9.20 & +29 & CP & p & 97.8059 & 19 & 3.55 \\
         & CP$-$56\,6334 & 104 & 9.90 & & B8\,V \\
         & HD 127817 & 81 & 9.15 & & B7\,IV \\
\smallskip
         & HD 127835 & 97 & 9.40 & & B7\,V \\
NGC 6281 & HD 153947 & 9 & 8.80 & & CP & n & 1395.1398 & 16 & 2.6841 & 2.662\tablefootmark{4} \\
         & HD 153948 & 15 & 9.35 & +40 & CP & y & 1395.1359 & 16 & 1.64 & 1.6370\tablefootmark{2} \\
         & BD$-$37\,11237 & 6 & 8.60 & & B9 \\
\smallskip
         & BD$-$37\,11241 & 5 & 8.30 & & A2 \\
NGC 6405 & CD$-$32\,13119 & 7 & 10.95 & +33 & CP & p & 6.0188 & 6 & data too poor \\
         & HD 318100 & 19 & 8.82 & +67 & CP & y & 1819.0577 & 13 & data too poor & 1.03012\tablefootmark{2} \\
         & HD 318107 & 77 & 9.35 & +94 & CP & y & 1819.0681 & 13 & data too poor & 4.527\tablefootmark{2} \\ 
         &           &    &      &     &    &   &           &    &               & 9.7085\tablefootmark{5} \\
         &           &    &      &     &    &   &           &    &               & 52.6\tablefootmark{2} \\
         & HD 160260 & 70 & 8.35 & & B9 \\
\smallskip
         & HD 160335 & 2 & 7.30 & & B3\,V \\
NGC 6475 & HD 162305 & 14 & 7.80 & +16 & CP & y & 1820.0393 & 14 & variable \\
         & HD 320764 & 23 & 8.90 & +20 & CP & y & 1820.0480 & 14 & variable? \\
         & HD 162515 & 42 & 6.52 & & B9\,V \\
\hline
\end{tabular}
\tablefoot{The primary targets are denoted with a spectral class
``CP''. The data for the $V$ magnitude, the spectral 
classification of the comparison stars, and
$\Delta$a measurements were taken from WEBDA 
(number W); the membership flag as determined in Sect. \ref{membs}. In addition, the results of our time 
series analysis with the time base of the
observations ($\Delta t$) and the number of data points (Num) for Str{\"o}mgren
$uvby$ are listed. The derived periods are significant
to the last listed digit. \\
\tablefoottext{a}{Hipparcos data available; results are discussed in Sect. \ref{hip}.} \\
\tablefoottext{b}{Our data set shows also a longer period which can not be resolved.} \\
\tablefoottext{1}{\citet{War04}} 
\tablefoottext{2}{\citet{North87}}
\tablefoottext{3}{\citet{Koen02}}
\tablefoottext{4}{\citet{North88}}
\tablefoottext{5}{\citet{Manf00}}
}
\label{list_stars}
\end{center}
\end{table*}

\section{Introduction}

The group of chemically peculiar (CP) stars on the upper main 
sequence display peculiar lines and line strengths, in addition to 
other peculiar features such as a strong global 
stellar magnetic field \citep{Bab47}. 
This subclass of B to F-type stars is characterized by 
variable line strengths and radial velocity changes as well as photometric 
variability of in general the same periodicity. 
One can usually distinguish between He-weak/strong, HgMn, Si, SrCrEu, and Am 
stars. The subgroup of SrCrEu
objects, for example, typically have overabundances of up to several dex for e.g., Sr, Cr, 
Eu, and other rare earth elements compared to the Sun.

Photometric variability of the CP star $\alpha^{2}$ CVn was first reported 
by \citet{Guth14}. The light curves can be fitted well
by a sine wave and its first harmonic with varying amplitudes for different 
photometric filter systems. For some CP stars, a double-wave structure of
the photometric light curves is detected \citep{Mai80}. However, similar 
magnetic field modulus variations are rare exceptions \citep{Mathy97}.

The variability
of CP stars is explained in terms of the
oblique rotator model \citep{Stibbs50}, according to which, 
the period of the observed light, spectrum, and magnetic field 
variations is the rotational period. Accurate
knowledge of the period of variability and its evolution in time for CP stars is
a fundamental step in understanding their complex behaviour,
especially as far as it concerns the phase relation
between the magnetic, spectral, and light variations \citep{Miku10}.

In 2004, an on-line catalogue of photometric observations of magnetic CP stars 
(mCPod\footnote{http://astro.physics.muni.cz/mcpod/}) was initiated 
\citep{Miku07}. This was intended to gather all 
available photometric data for CP stars, into a single, freely accessible 
database.
The archive presently contains about 150\,000 individual measurements of 151 CP 
stars and is being constantly updated with the latest photometric data.

In this paper, we present Str{\"o}mgren $uvby$ time series for 27 CP
stars within the boundaries of open clusters. In addition, we queried the
Hipparcos photometric database \citep{Leeuw97} for entries of
our targets and found four matches. Our observations cover a time interval of
about six years with typically fifteen measurements for each object.
A detailed time series analysis was performed, implementing five different
methods to minimize spurious detections. 

Because the variability is assumed to be related to the rotation, a detailed study of
the light curves is essential, not only to determine astrophysical
parameters using more realistic stellar model atmospheres, but also to interpret the substantial 
information provided when mapping the stellar surface with Doppler imaging techniques
\citep{Lehm07}. The choice of open cluster CP stars allows us to establish  
fundamental parameters such as the age, distance, reddening, and metallicity  
of numerous cluster members more accurately than 
usually possible for Galactic field stars.

\begin{table}
\begin{center}
\caption[]{Log of observations with the extinction coefficients
for the Str{\"o}mgren $uvby$ filters. For the majority of the
observations, the standard extinction coefficients for LaSilla
fit the data very well.}
\begin{tabular}{llccccc}
\hline
\hline
Teles. & Obs. & Night & $k_u$ & $k_v$ & $k_b$ & $k_y$ \\
          &          & [JD]   & [mag] & [mag] & [mag] & [mag] \\
\hline
Bochum & HMM & 2446581 & 0.570 & 0.350 & 0.230 & 0.170 \\
       &               & 2446582 & 0.570 & 0.350 & 0.230 & 0.170 \\
       &               & 2446583 & 0.570 & 0.350 & 0.230 & 0.170 \\
       &               & 2446584 & 0.570 & 0.350 & 0.230 & 0.170 \\
       &               & 2446585 & 0.570 & 0.350 & 0.230 & 0.170 \\
       &               & 2446587 & 0.570 & 0.350 & 0.230 & 0.170 \\
       &               & 2447963 & 0.570 & 0.350 & 0.230 & 0.170 \\
       &               & 2447964 & 0.570 & 0.350 & 0.230 & 0.170 \\
       &               & 2447965 & 0.570 & 0.350 & 0.230 & 0.170 \\
       &               & 2447966 & 0.570 & 0.350 & 0.230 & 0.170 \\
       &               & 2447967 & 0.570 & 0.350 & 0.230 & 0.170 \\
       &               & 2447968 & 0.570 & 0.350 & 0.230 & 0.170 \\
       &               & 2447969 & 0.570 & 0.350 & 0.230 & 0.170 \\
       &               & 2447970 & 0.570 & 0.350 & 0.230 & 0.170 \\
       &               & 2447971 & 0.570 & 0.350 & 0.230 & 0.170 \\
       &               & 2447972 & 0.570 & 0.350 & 0.230 & 0.170 \\
       &               & 2447973 & 0.570 & 0.350 & 0.230 & 0.170 \\
       &               & 2447974 & 0.570 & 0.350 & 0.230 & 0.170 \\
       &               & 2447976 & 0.570 & 0.350 & 0.230 & 0.170 \\
SAT & CT & 2448307 & 0.570 & 0.350 & 0.230 & 0.170 \\
    &            & 2448308 & 0.570 & 0.350 & 0.230 & 0.170 \\    
    &            & 2448309 & 0.570 & 0.350 & 0.230 & 0.170 \\
    &            & 2448310 & 0.570 & 0.350 & 0.230 & 0.170 \\
    &            & 2448311 & 0.570 & 0.350 & 0.230 & 0.170 \\
    &            & 2448313 & 0.570 & 0.350 & 0.230 & 0.170 \\
    &            & 2448314 & 0.570 & 0.350 & 0.230 & 0.170 \\
    &            & 2448315 & 0.570 & 0.350 & 0.230 & 0.170 \\
    &            & 2448308 & 0.570 & 0.350 & 0.230 & 0.170 \\
    &            & 2448308 & 0.570 & 0.350 & 0.230 & 0.170 \\   
SAT & HH & 2448392 & 0.570 & 0.350 & 0.230 & 0.170 \\
    &              & 2448393 & 0.570 & 0.350 & 0.230 & 0.170 \\
    &              & 2448394 & 0.570 & 0.350 & 0.230 & 0.170 \\
    &              & 2448398 & 0.570 & 0.350 & 0.230 & 0.170 \\
    &              & 2448400 & 0.570 & 0.350 & 0.230 & 0.170 \\
    &              & 2448405 & 0.570 & 0.350 & 0.230 & 0.170 \\
		&              & 2448407 & 0.592 & 0.369 & 0.244 & 0.184 \\
    &              & 2448408 & 0.592 & 0.369 & 0.244 & 0.184 \\
    &              & 2448409 & 0.592 & 0.369 & 0.244 & 0.184 \\
    &              & 2448410 & 0.592 & 0.369 & 0.244 & 0.184 \\
    &              & 2448411 & 0.735 & 0.455 & 0.320 & 0.240 \\
    &              & 2448413 & 0.575 & 0.355 & 0.230 & 0.170 \\
    &              & 2448414 & 0.575 & 0.355 & 0.230 & 0.170 \\
    &              & 2448415 & 0.575 & 0.355 & 0.230 & 0.170 \\
    &              & 2448416 & 0.575 & 0.355 & 0.230 & 0.170 \\ 
SAT & HMM & 2448638 & 0.720 & 0.480 & 0.350 & 0.270 \\
    &               & 2448639 & 0.660 & 0.420 & 0.305 & 0.240 \\
    &               & 2448640 & 0.630 & 0.380 & 0.280 & 0.230 \\
    &               & 2448641 & 0.590 & 0.360 & 0.250 & 0.200 \\
    &               & 2448642 & 0.600 & 0.355 & 0.250 & 0.190 \\
    &               & 2448643 & 0.600 & 0.360 & 0.255 & 0.205 \\
    &               & 2448644 & 0.590 & 0.350 & 0.240 & 0.190 \\
\hline
\end{tabular}
\label{obs_log}
\end{center}
\end{table}

\section{Target selection, observations, and reduction}

This work was initiated by the European working group on chemically peculiar 
stars of the upper main sequence \citep{Mathy89}. It continues and complements
an extensive survey performed by \citet{North92}. A continuation of these
efforts had already been initialized using the Rapid Eye
Mount (REM) telescope at La Silla (P.I.: L.~Fossati). 

The targets are established
as candidate CP stars in open clusters mainly by the extensive photoelectric
$\Delta$a survey (e.g. \citealt{Mai93}). The comparison stars were chosen in the vicinity of the
targets within the same magnitude range (if possible). Table \ref{list_stars} lists the 
target and comparison stars and their WEBDA\footnote{http://www.univie.ac.at/webda} 
numbers. The data for the $V$ magnitude, spectral type, and $\Delta$a measurement \citep{Mai76} 
were also taken from WEBDA.

The star CP$-$60\,944 is a close visual binary with almost equally bright
components. In the literature, there is a constant confusion about the designation
of this object. \citet{Rens09} list two entries in their most recent
catalogue of Am and Ap stars, with $V$-magnitudes of 8.3 and 8.8, respectively.
Both components are classified as B8\,Si stars therein. 
The published $\Delta$a value \citep{MH81} is a combined 
value of both components, since it was impossible to separate them.

In addition, we queried the Hipparcos photometric database \citep{Leeuw97} for entries of the 
photoelectrically observed targets. In total, we found four matching entries: HD~56343, HD~87405, 
HD~89856, and HD~127924. 

The photometric observations in the Str{\"o}mgren $uvby$
system were performed from 1986 to 1992 at the 50\,cm Danish
Telescope (SAT) and with the 61\,cm Bochum telescope at LaSilla.
The Bochum telescope was equipped with a dry-ice-cooled EMI 9502 
photomultiplier, whereas the SAT contains an EMI 9789QB one. A detailed
log of observations is given in Table \ref{obs_log}.

All data were corrected for atmospheric extinction
and sky background. In a first step, the standard extinction
coefficients for LaSilla of the individual filters were used.
For most of the nights, this choice was valid and 
very effective. For the observing nights in 1991 and 1992,
the Volcano eruption of Pinatubo, we had to recalculate
the coefficients as listed in Table \ref{obs_log}. Finally,
all observations were corrected to a heliocentric standard time frame.

All photometric measurements are available in electronic form 
at the CDS, from the first author and will be included in mCPod.

\section{Time series analysis}

The time series analysis of unevenly spaced data for a
wide variety of time frames is not straightforward. Several
effects (for example aliasing) severely influence the corresponding 
diagnostic diagrams.

To minimize the effects introduced by the limitations of
the data sets and determine a stable 
solution, we applied five different time series analysis methods:
\begin{itemize}
\item Lafler - Kinman method \citep{Lafl76};
\item Modified Lafler - Kinman method \citep{Hensb77};
\item Phase-dispersion Method \citep{Stell78};
\item Renson method \citep{Rens78};
\item Standard Fourier method \citep{Deem75}.
\end{itemize}
The first of the four aforementioned methods are optimized to
analyse sparsely distributed data sets. The results of the different
methods were compared and the significant periods extracted. The
complete set of results and published ones are
listed in Table \ref{list_stars}. The derived periods are significant
to the last given digit. All computations were performed using the 
programme package Peranso\footnote{http://www.peranso.com/}.

The decision about the (non-)variability of the individual objects
was made after applying the above-mentioned time-series analysis
methods. If the detection limit of all four algorithms failed, we assigned
an object ``constant'' within the precision of our measurements. The definition
of these limits can be found in the specific references and are mainly
based on the appearance of significant peaks in the amplitude-frequency
domain.

\begin{figure*}
\begin{center}
\includegraphics[width=160mm]{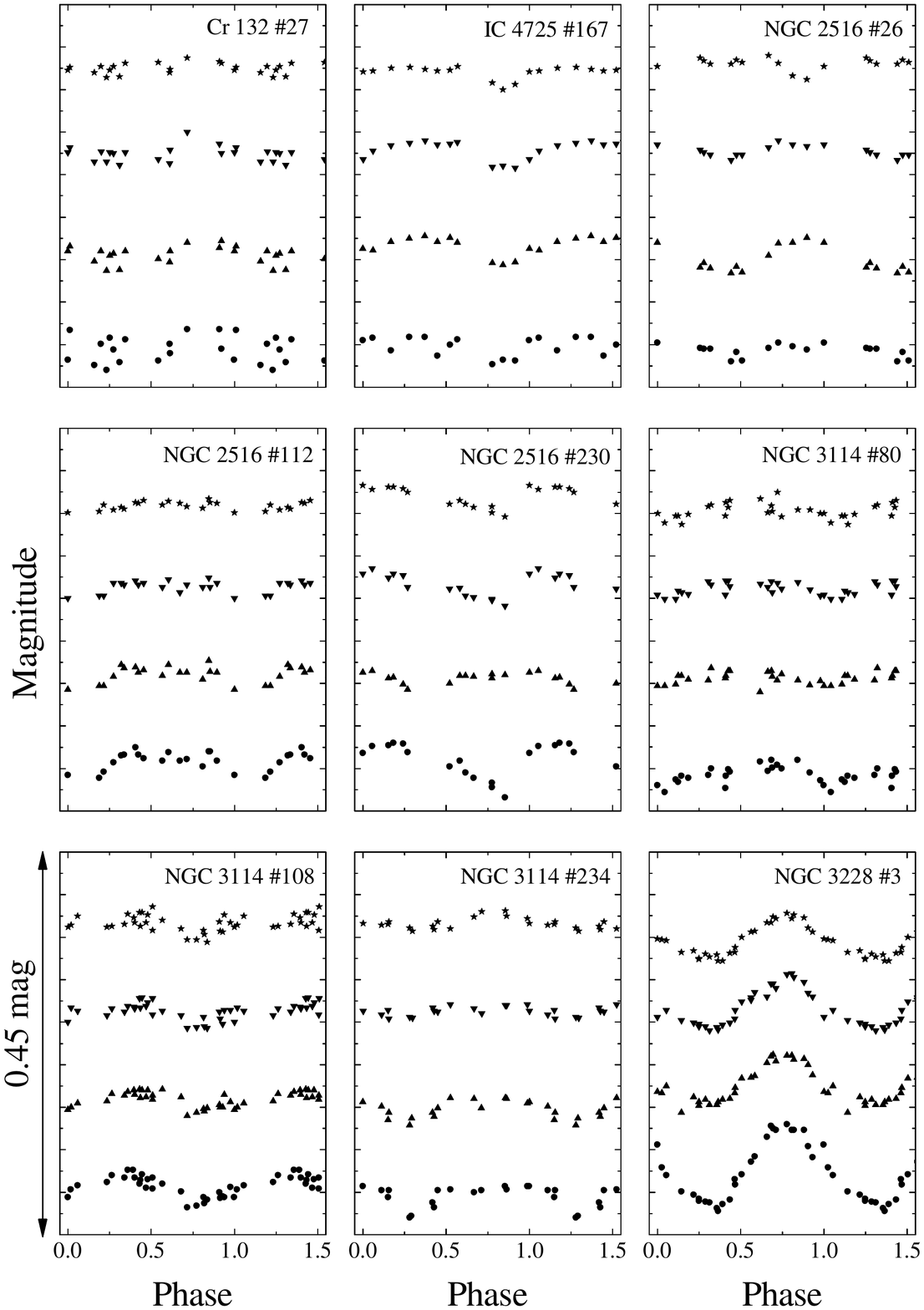}
\caption[]{Phase diagrams of differential Str{\"o}mgren $u$ (circles),
$v$ (upper triangles), $b$ (lower triangles), and $y$ (star symbols) magnitudes
for the variable program stars with well-established periods. The interval between
two major ticks is 0.05\,mag.}
\label{phase1}
\end{center}
\end{figure*}

\begin{figure*}
\begin{center}
\includegraphics[width=160mm]{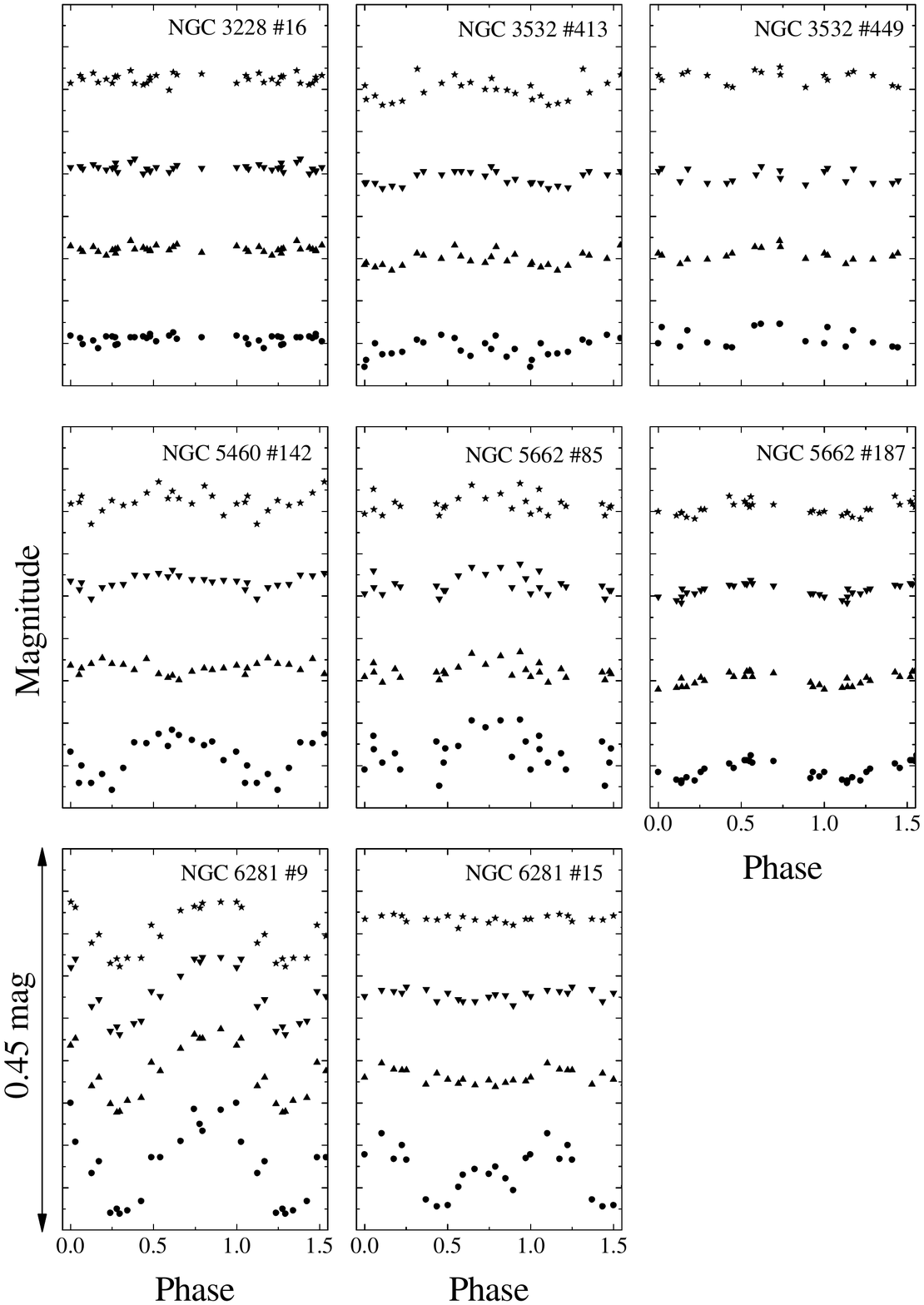}
\addtocounter{figure}{-1}
\caption[]{continued}
\label{phase2}
\end{center}
\end{figure*}

\begin{table}
\begin{center}
\caption[]{The observed open clusters with their parameters
based on the list by \citet{PN06}.}
\begin{tabular}{lclcr}
\hline
\hline
Cluster & Name &  \multicolumn{1}{c}{$D$} & $E(B-V)$ & \multicolumn{1}{c}{Age} \\
 & & \multicolumn{1}{c}{[pc]} & \multicolumn{1}{c}{[mag]} & \multicolumn{1}{c}{[Myr]} \\
\hline
Cr 132   & C0712-310 & 438(39) & 0.03(1) &  23(10) \\
IC 4725  & C1828-192 & 615(65) & 0.48(2) &  75(15) \\
NGC 2516 & C0757-607 & 402(32) & 0.10(4) & 141(21) \\
NGC 3114 & C1001-598 & 914(19) & 0.06(3) & 128(38) \\
NGC 3228 & C1019-514 & 511(22) & 0.02(1) & 102(17) \\
NGC 3532 & C1104-584 & 492(8)  & 0.04(1) & 262(46) \\ 
NGC 5460 & C1404-480 & 673(76) & 0.12(2) & 167(54) \\
NGC 5662 & C1431-563 & 684(60) & 0.32(1) &  77(20) \\ 
NGC 6281 & C1701-378 & 516(35) & 0.15(1) & 285(58) \\
NGC 6405 & C1736-321 & 473(16) & 0.14(2) &  71(21) \\ 
NGC 6475 & C1750-348 & 258(21) & 0.07(3) & 267(62) \\
\hline
\end{tabular}
\label{list_clusters}
\end{center}
\end{table}

Figure \ref{phase1} shows the phase diagrams of all programme stars
with a well-established period in differential Str{\"o}mgren $uvby$. For the
folding process, the periods listed in Table \ref{list_stars} are used.
The light curves of almost all targets are in phase for
the four used filters. However, the $y$ data for NGC 2516 \#26 and NGC 3532 \#449, as well as the 
$v$ data for NGC 5460 \#142 are not in phase with those of the other filters. The
light curves of NGC 2516 \#230 and NGC 6281 \#15 are very asymmetric, especially
for the $u$ filter. The shape of the light curve for NGC 6281 \#15 
is indicative of a double wave. These two stars certainly require further attention 
and complementary data to shed more light on the cause of this asymmetry.

For NGC 3114 \#80 and NGC 3532 \#413, we find at least two periods for each star
on widely different timescales. With the given data set, we are unable to decide
which period is due to classical rotation. In the phase diagrams of these two objects, the 
long-term trend was subtracted from the data. These two stars may be similar in some
ways to HR 1297, a B9\,Si star, for which periods of 1.06457\,days and
15.749\,days were reported by \citet{Adm00}. The astrophysical reason for this
behaviour remains unknown but it is certainly connected to the highly variable properties
of the stellar photosphere.

For NGC 3228 \#16 we used the
period of 0.84\,days, which fits the data more closely than the longer period
of 0.94\,days.

For NGC 2516 \#230, \citet{War04} presented $I$ photometry and inferred a
period of 1.943\,days from a Fourier fit. This finding agrees with our result.

By combining the data for NGC 3114 \#234 with those published by \citet{North87}, 
we obtain a more accurate measurement of the period of 2.31808\,days.
The shape of the Geneva photometry light curves shown in
\citet{North87} are very similar to ours. When combining the data sets, we used a mean
magnitude calculated from all seven filters as well as the individual  
ones. The deduced period for these different procedures is numerically the same up to the fifth
digit. Our light curves do not show as prominent a secondary minimum as those from
the Geneva data. However, we are unable to definitely rule out a double wave 
signature, especially for Str{\"o}mgren $y$.

For the three CP stars in NGC~6405, the data are of too poor quality to draw any
definitive conclusion. Even the combination with previously published photometry
does not result in any significant improvement of the situation. However,
since our photometric data will be publicly available, they are an important 
contribution to long-term studies of this star group.

Four stars of our sample, namely IC 4725 \#98, IC 4725 \#153, NGC 3114 \#25,
NGC 6475 \#14, and NGC 6475 \#23, are certainly variable according to the 
detection limits of the applied time-series methods, but we are unable to deduce reliable
periods from our data sets. These objects are worthwhile targets for follow-up
observations. The time series analysis of our data set for NGC 6475 \#23 is rather inconclusive.
From the different results and an inspection of the light curve by eye, we 
conclude that this star is indeed variable. 

In Fig. \ref{phase2}, the phase diagrams of all stars except for IC 4725 \#98,
which has few data points, are plotted. For each object we took the period derived
for the differential Str{\"o}mgren $v$ magnitudes. The range of calculated periods agrees well
with those of the other program stars.

\begin{figure}
\begin{center}
\includegraphics[width=85mm]{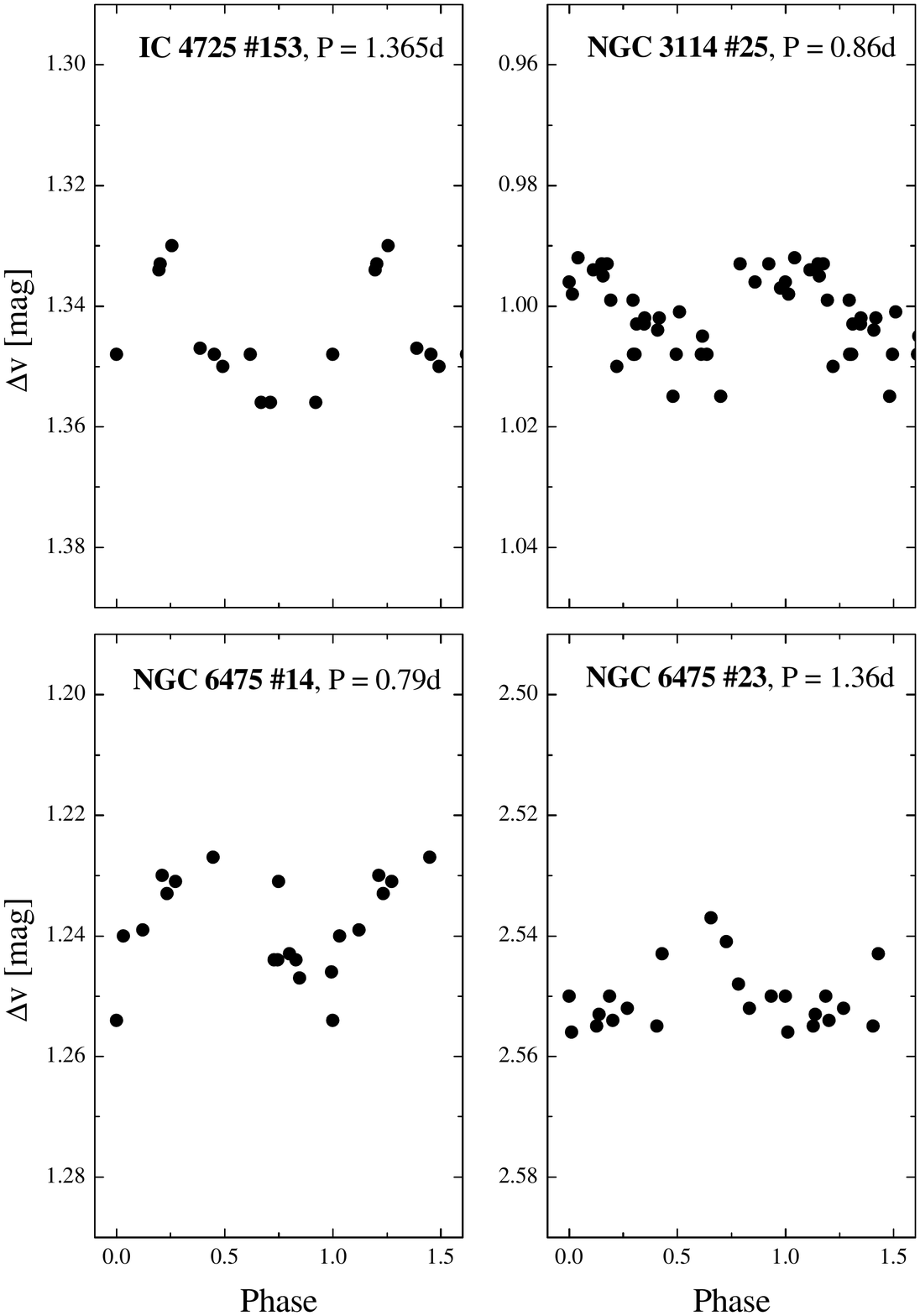}
\caption[]{Phase diagrams of differential Str{\"o}mgren $v$ magnitudes
for the variable program stars with an uncertain period.}
\label{phase2}
\end{center}
\end{figure}

\section{Hipparcos photometry} \label{hip}

For four target stars, we retrieved Hipparcos data and analysed them together with
the ground-based data, i.e. Cr 132 \#27 (HIC 120047), NGC 3114 \#108 (HIC 49218), 
NGC 3228 \#3 (HIC 50677) and NGC 5662 \#187 (HIC 71398). Among these stars, only
NGC 3228 \#3 was flagged as variable from the Hipparcos data.

The data sets cover a typical time frame of three years with
about 130 measurements of each star.

Following the advice given in the catalogue publication,
measurements derived from data with identified reduction problems were
not considered. Since the Hipparcos data have uncertainties
of between 1\% and 2\% at the involved magnitudes, comparable to
the expected amplitudes of the light variation, a weighted
average of data obtained over a small section of the same
rotation cycle, typically up to $0.05 P_{rot}$, was used.
Indeed, a smaller amount of more accurate observations is
more helpful to period searches and Hipparcos observations
tend to be clustered in time because of the observing procedure (two
observations within 20 minutes, repeated commonly two hours
later).

The time series analysis was performed
such that we attempted to identify the optimal period by fitting to all quantities $m$
(all individual Str{\"o}mgren filters and $H_p$) a function
\begin{equation}
m = m_0 + A (\cos 2\pi (\phi - \phi_0) + B \cos 4\pi (\phi - \phi_1) ),
\end{equation}
where $m_0$, $A$, $B$, $\phi_0$ and $\phi_1$ are free parameters. When the shape 
of the light curves did not significantly differ in the filters, we took
the same values of $B$, $\phi_0$, and $\phi_1$ in all channels, only varying
$m_0$ and $A$ from filter to filter.

{\bf Cr 132 \#27:} This is the faintest Hipparcos target of our sample. The
analysis of the Hipparcos data alone results in a period that
is twice as long as deduced from the $uvby$ photometry. This may be indicative of
an apparent double wave. However, merging the two data sets
does not help us to prove with this interpretation. 

{\bf NGC 3114 \#108:} From the $uvby$ photometry we derive a period of
2.113\,days. Adding the Hipparcos measurements, the range of possible
periods can be narrowed down to 2.1250\,$-$\,2.1265\,days. This
interval contains two solutions with overlapping significance
intervals at the 1.5$\sigma$ level. The highest probability occurs at
2.1253 and at 2.1262\,days, respectively. 

{\bf NGC 3228 \#3:} The $uvby$ light curves deduced for 
this object are rather noisy, but the derived period is 
consistent with that derived from the Hipparcos observations and is compatible
with the result obtained by \citet{Koen02} from the Hipparcos data.

{\bf NGC 5662 \#187:} In spite of the low amplitude, the periodicity 
is still detectable from an analysis of the Hipparcos magnitudes. 
It is most clearly visualized by forming weighted averages over individual 
measurements that were obtained within one fifteenth of the same 
stellar rotation cycle. The combination of both $uvby$ and Hipparcos 
photometry allows us to deduce the period given in Table \ref{list_stars}.

\section{Membership}\label{membs}

In more in-depth investigations (e.g., of evolutionary aspects), it is essential to know whether the respective object 
belongs to the cluster. Several criteria can be used to determine membership. Apart from 
photometric criteria, the most commonly used are certainly based on proper motion. 
These, and both radial velocities and parallaxes would help assign membership, but 
data of the latter two in particular are unfortunately sparsely 
available for our programme objects. Several comprehensive studies are available, which already provide 
membership probabilities in open clusters using proper motion \citep{Baumg00, Dias01, Dias06, Khar04}, 
from which we compiled their results as well as their respective data source. Furthermore, we considered the results 
of \citet{Javak06}, which are based on the same data (Tycho-2) as \citet{Dias01}, but a  
different method was applied to obtain membership probabilities. These datasets were complemented with  
data from other previous studies \citep{Const69, Franc89, King78a, King78b, King80} for 
NGC~6475, IC~4725, NGC~3532, NGC~2516, and NGC~5662, respectively.

Since slightly different cluster means 
were determined in the various studies and also different scales were used, the use of probability values 
of a single source can lead to wrong conclusions. They have to be compared with other investigations, 
including those examining the respective source data and the deviation from the cluster mean motion. A 
demonstrative example are the results for HD~127924, for which \citet{Baumg00} list 9\%, although this star 
deviates 2.6\,mas only from the cluster mean. The same value was obtained by
\citet{Dias06} with a deviation of 7.8\,mas, but their cluster mean differs significantly from all other studies.

We therefore followed the approach given in 
\citet{Land07} to divide membership into y (member), p (probable member), ? (probably non-member), and n (non-member).
We refer to \citet{Land07} for a more detailed explanation of the problems connected with membership probabilities. 
We queried the latest release of UCAC3 \citep{Zach09} to check the consistency of the proper motions and 
assigned a membership flag to the programme objects based on this property. As in \citet{Land07}, we performed a similar
assignment
for the position of the objects in the Hertzsprung-Russell diagram (HRD) relative to the isochrone of the respective 
cluster age. Furthermore, parallaxes were extracted from the new Hipparcos reduction by \citet{Leeuw07} and compared with 
the cluster distances. As a fourth membership parameter, we used radial velocities taken from \citet{Ami81}, \citet{Franc89}, 
\citet{Gies85}, \citet{Gonz00}, \citet{Gonz01}, \citet{Land08}, as well as \citet{Luck00}, 
which were compared with both the cluster mean of the respective sources and 
those of \citet{Mer08}. However, these results have to be evaluated with caution, if only a few measurements are 
available since binarity cannot be excluded. Using all available information, a final membership flag 
was assigned, which is given in Table \ref{membership} together with the result by \citet{Land07}. In the following, 
non-members, with results that differ significantly from both \citet{Land07} and other results are discussed.

Three stars (HD~89856, 96729, and 153947) that deviate strongly in the HRD: only unusually large 
errors in temperature of the order of about 2000\,K would put them close to the corresponding isochrone. All other available 
membership parameters are also uncertain, so they were classified as non-members in agreement with \citet{Land07}. Furthermore, 
the position of HD~153947, not included in this reference, is located about four times the cluster 
radius (4$\arcmin$, \citealt{Dias02}) away from the cluster center. 

HD~65712 in NGC~2516 is also located far from the cluster center, at a distance of
about three times the cluster radius of 15$\arcmin$ \citep{Dias02}. 
Since the two available parameters are indicative of membership, and the star is 
just within the much larger radius determined by  \citet{Khar04}, a probable 
membership was deduced.

For the star HD~87240 in NGC~3114, its
radial velocity data and the position in the HRD support membership, in contrast to most proper motions. However, the data of 
UCAC3, which deviates strongly from all others, do not contradict membership. Therefore, a probable membership was concluded 
for this star. Several membership criteria are available for HD~87405 in the same cluster. The radial velocities of 
\citet{Gonz01} exclude membership, listing a probability of 0\%. In contrast, the results by \citet{Ami81} are in 
good agreement with their measurement of cluster mean. Both presented several measurements that provided no
evidence of binarity. We prefer the 
results of the first of these aforementioned studies, since they are based on higher resolution CCD measurements than the photographic 
plates used by \citet{Ami81}. This star was therefore classified as a probable non-member as in \citet{Land07}. 

According 
to its proper motion and all available studies, the star CP$-$56 6330 cannot be a member of the cluster NGC~5662, although its position 
in the HRD suggests the contrary. The final classification as a non-member agrees with \citet{Land07}. 
Its proper motion as well as its HRD position fully support membership of HD~56343 to Collinder~132, but its parallax 
(0.26 $\pm$ 1.1) would place it as a background object, which also holds for the ``original'' Hipparcos data (0.66 $\pm$ 1.09). 
However, we note that the nature of this open cluster continues to be discussed in the literature 
\citep{Clar77, Baumg98, Cab08}. Therefore, we are not able to draw a final membership conclusion for this star. For the 
visual binary CP$-$60 944, both components are at least probable members, whereas the A component is according to its proper 
motion and position in the HRD a definite member (see also \citealt{Land07}).

\begin{table}
\caption{Membership of the programme objects.}
\label{membership} 
\centering 
\renewcommand{\footnoterule}{}
\begin{tabular}{l l l l } 
\hline\hline 
Cluster & CP Star & \multicolumn{2}{c}{Membership}   \\
        &         & $\mu$/$\nu$/ph/$\pi$ & final \\  
\hline 
Cr 132   & HD 56343      & y/$-$/y/n    & $-$ (p)\tablefootmark{a}   \\
IC 4725  & BD$-$19 5044L & p/$-$/p/$-$  & p (p)   \\
         & HD 170836     & y/y/p/$-$    & y (p)   \\
         & HD 170860     & p/n\tablefootmark{b}/y/$-$    & p (p)   \\
NGC 2516 & CP$-$60 944   & y/$-$/y/$-$  & y (y)\tablefootmark{a}   \\
         & HD 65712      & y/$-$/y/$-$  & p (y)\tablefootmark{a}   \\
         & HD 66295      & y/y/y/$-$    & y (y)   \\
NGC 3114 & HD 87240      & p/y/y/$-$    & p (n)\tablefootmark{a}   \\
				 & HD 87405      & p/n/y/p      & ? (?)\tablefootmark{a}   \\
				 & HD 87752      & p/y/y/$-$    & y ($-$)   \\
				 & HD 304841     & p/$-$/y/$-$  & p (p)   \\
				 & HD 304842     & p/y/y/$-$    & y (p)   \\
NGC 3228 & HD 89856      & p/$-$/n/p    & n (n)\tablefootmark{a}   \\
				 & HD 298053     & p/$-$/y/$-$  & p (p)   \\
NGC 3532 & HD 96040      & p/$-$/p/$-$  & p (p)   \\
				 & HD 96729      & p/$-$/n/$-$  & n (?)\tablefootmark{a}   \\
NGC 5460 & HD 122983     & y/$-$/p/$-$  & p (p)   \\
				 & HD 123225     & y/$-$/y/$-$  & y (y)   \\
NGC 5662 & CP$-$56 6330  & n/$-$/y/$-$  & n (n)\tablefootmark{a}   \\
			   & HD 127924     & y/$-$/p/p    & p (y)   \\
NGC 6281 & HD 153947     & p/$-$/n/$-$  & n ($-$)\tablefootmark{a}   \\
				 & HD 153948     & y/p/y/$-$    & y (y)   \\
NGC 6405 & CD$-$32 13119 & p/$-$/y/$-$  & p (p)	  \\			 
         & HD 318100     & y/y/y/$-$    & y (y)   \\
			   & HD 318107     & y/$-$/y/$-$  & y (y)  \\
NGC 6475 & HD 162305     & y/y/y/$-$    & y (y)  \\
         & HD 320764		 & y/y/y/$-$    & y (p)  \\   				 			   				 				 				 
				 				           
\hline 
\end{tabular}
\tablefoot{The final membership was deduced from individual ones based on proper motion ($\mu$), radial velocity ($\nu$), 
position in the HRD (ph), and parallaxes ($\pi$). In parenthesis, the membership obtained by 
\citet{Land07} is given. \\
\tablefoottext{a}{Membership discussed in the text.}\\
\tablefoottext{b}{Only a single measurement by \citet{Land08}; binarity cannot be excluded.}
}
\end{table}

\begin{table}
\caption{Derived logarithmic luminosities and effective temperatures of our programme stars.}
\label{logs} 
\centering 
\begin{tabular}{lcc} 
\hline\hline 
CP Star & log\,$T_{\rm eff}$ & log\,$L/L_{\sun}$ \\
\hline
BD $-$19 5044L	&	4.094	&	2.25 \\
CD $-$32 13119	&	3.914	&	1.05 \\
CP $-$56 6330	&	4.088	&	1.95 \\
CP $-$60 944	&	4.095	&	2.24 \\
HD 56343	&	4.063	&	1.70 \\
HD 65712	&	4.000	&	1.52 \\
HD 66295	&	4.043	&	1.75 \\
HD 87240	&	4.092	&	2.29 \\
HD 87405	&	4.092	&	2.74 \\
HD 87752	&	4.105	&	2.28 \\
HD 89856	&	4.165	&	2.16 \\
HD 96040	&	4.026	&	1.42 \\
HD 96729	&	4.083	&	1.57 \\
HD 122983	&	4.032	&	1.89 \\
HD 123225	&	4.084	&	2.40 \\
HD 127924	&	4.113	&	2.50 \\
HD 153947	&	4.116	&	2.29 \\
HD 153948	&	4.026	&	1.84 \\
HD 162305	&	4.008	&	1.74 \\
HD 170836	&	4.162	&	2.89 \\
HD 170860\,A	&	4.137	&	2.63 \\
HD 298053	&	3.936	&	1.09 \\
HD 304841	&	4.089	&	2.15 \\
HD 304842	&	4.094	&	2.25 \\
HD 318100	&	4.025	&	1.60 \\
HD 318107	&	4.070	&	1.90 \\
HD 320764	&	3.932	&	1.26 \\
\hline 
\end{tabular}
\end{table}

\begin{figure}
\begin{center}
\includegraphics[width=85mm]{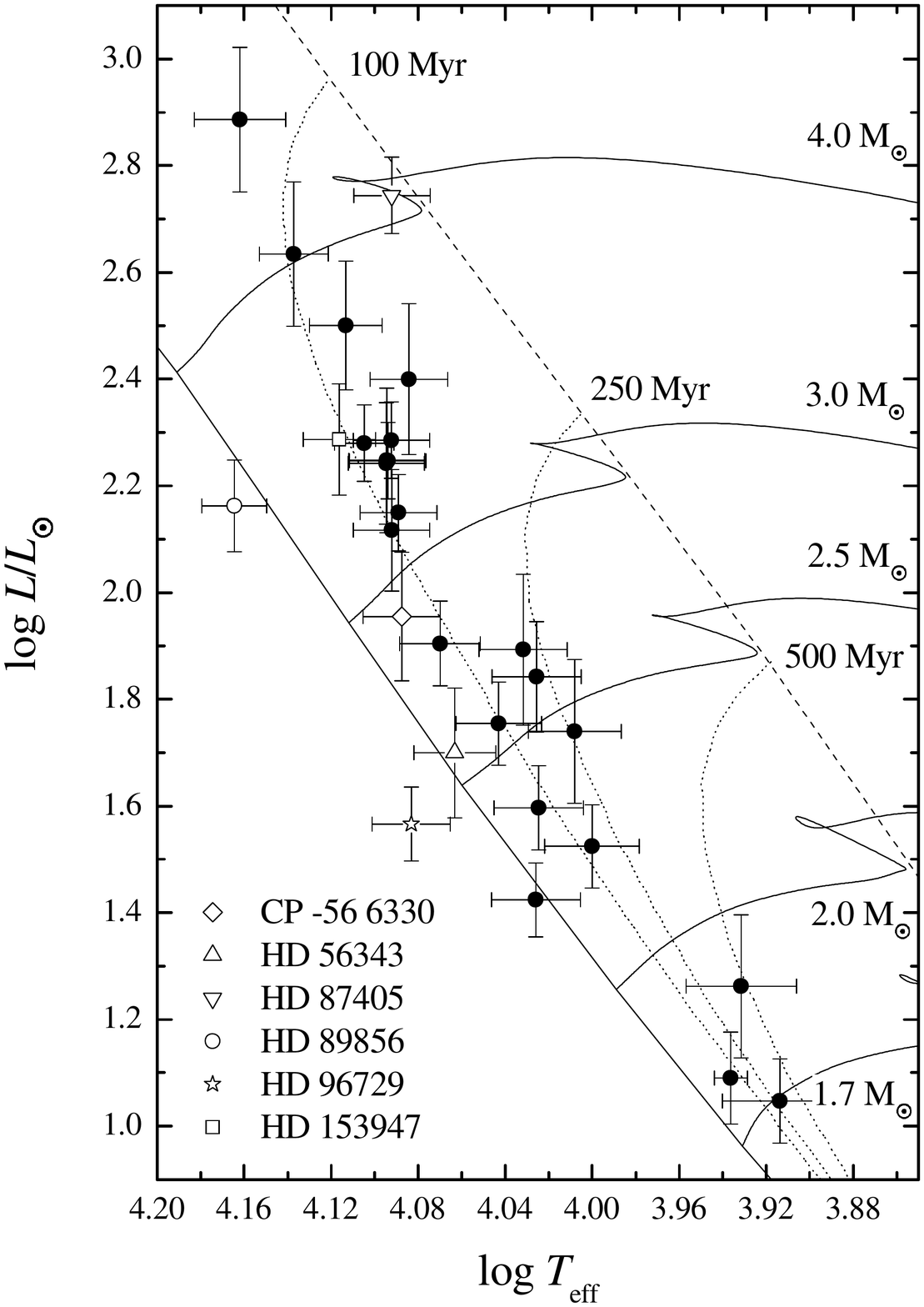}
\caption{The location of the investigated CP stars in a log\,$L/L_{\sun}$ versus
log\,$T_{\rm eff}$ diagram. Filled symbols denote members, whereas open
symbols are non-members. The dashed line denotes the 
terminal-age main-sequence; the evolutionary tracks for individual masses are interpolated
within the solar metallicity ones listed by \citet{Schall92}, \citet{Schaer93a}, \citet{Schaer93b}, and
\citet{Charb93}.}
\label{hrd}
\end{center}
\end{figure}

\section{Results and conclusions}

We have investigated photometrically a total of 27 CP stars within the boundaries of
open clusters.

We established variability for fourteen CP stars with previously unknown
rotation periods, and confirmed the rotation periods of eight more stars with
an increase in precision for two of them.

The applied open cluster parameters (Table \ref{list_clusters})
are averaged ones based on a set of literature values.
Starting with the comprehensive compilation by \citet{PN06}, we searched the 
literature for additional (new or overseen) parameters. They were all checked for plausibility 
by using appropriate isochrones and available photometric data taken from the WEBDA database. 
Significant different results were removed and finally a strict average and standard 
deviations were calculated. 

The log\,$T_{\rm eff}$ values were calculated using the calibration
of \citet{Net08} for CP stars.
We made use of the GCPD catalogue\footnote{http://obswww.unige.ch/gcpd/gcpd.html} 
and the WEBDA database to extract all
available photometric data of the programme stars. To obtain 
absolute magnitudes (assuming that all programme stars are indeed 
cluster members), we took averaged values and the mean cluster 
distances (see Table \ref{list_clusters}). Since the stars in NGC~2516 in particular,
exhibit a strong 
differential reddening, individual reddening values were determined whenever 
possible for all objects as suggested by \citet{Net08}. Within the 
latter reference a bolometric correction for magnetic CP stars was also
introduced, which was used to calculate the luminosity. For the 
remaining Am and HgMn objects, the bolometric correction by \citet{Bal94} 
for normal stars was applied. In Table \ref{logs}, the individual
values are listed.

Figure \ref{hrd} shows the location of the investigated CP stars in a 
log\,$L/L_{\sun}$ versus log\,$T_{\rm eff}$ diagram. The dashed line denotes the terminal-age
main-sequence. The evolutionary tracks for individual masses and ages are interpolated
between the solar metallicity ones listed by \citet{Schall92}, \citet{Schaer93a}, \citet{Schaer93b}, and
\citet{Charb93}.

Two stars, HD 89856 and
HD 96729 are located below the ZAMS and deviate significantly from the apparent
cluster age (Table \ref{list_clusters}). These objects are definite non-members according 
to our analysis and that of \citet{Land07}. We also marked the location of the other
questionable cluster members as discussed in Sect. \ref{membs}.

The target stars cover the typical mass range for mid B to late A type 
main-sequence objects from about 1.7M$_{\sun}$ to 4.5M$_{\sun}$ as other members of
this group \citep{Poehn05}. The photometric periods of 0.7--4.5~days are 
consistent with the typical rotation velocities of CP stars \citep{North92}.

There is a hint of the period decreasing with increasing age, in particular 
for stars with a mass below 3M$_{\sun}$, but this is not statistically significant
due to poor number statistics. The same is true for a possible correlation 
of the period with the stellar mass and effective temperature. However, with the
ongoing observations, more light will be shed on these important topics.

Our observations fill an important gap in previous photometric long-time studies of
CP stars. The apparent open cluster members are excellent targets for follow-up 
observations, based on for example polarimetry, high-resolution spectroscopy, and surface mapping 
techniques.

Follow-up observations within our framework are already under way with the Rapid Eye Mount
(REM) telescope at La Silla. This will help us to understand the
apparent stellar cycles, such as that of the Sun, for this group of magnetic peculiar objects.

\begin{acknowledgements}
We would like to thank Gautier Mathys for significantly improving this paper.
This work was supported by the financial contributions of the Austrian Agency for International 
Cooperation in Education and Research (WTZ CZ-11/2008, CZ-10/2010, and HR-14/2010), the City of Vienna 
(Hochschuljubil{\"a}umsstiftung project: H-1930/2008), the Forschungsstipendium der 
Universit{\"a}t Wien (F-416), a MOEL grant of the {\"O}FG (Project \#388) and a travel grant 
from the Swedish Research Council. 
UH acknowledges support from the Swedish National Space Board.
This research has made use of the WEBDA 
database, operated at the Institute for Astronomy of the University of Vienna. 
\end{acknowledgements}


\begin{thebibliography}{}
\bibitem[\protect\citeauthoryear{Adelman}{2000}]{Adm00} Adelman, S. J 2000, \aaps, 146, 13
\bibitem[\protect\citeauthoryear{Amieux \& Burnage}{1981}]{Ami81} Amieux, G., \& Burnage, R. 1981, \aaps, 44, 101
\bibitem[\protect\citeauthoryear{Babcock}{1947}]{Bab47} Babcock, H. W. 1947, \apj, 105, 105
\bibitem[\protect\citeauthoryear{Balona}{1994}]{Bal94} Balona, L. A. 1994, \mnras, 268, 119
\bibitem[\protect\citeauthoryear{Baumgardt}{1998}]{Baumg98} Baumgardt, H. 1998, \aap, 340, 402
\bibitem[\protect\citeauthoryear{Baumgardt et al.}{2000}]{Baumg00} Baumgardt, H., Dettbarn, C., \& Wielen, R. 2000, \aaps, 146, 251
\bibitem[\protect\citeauthoryear{Caballero \& Dinis}{2008}]{Cab08} Caballero, J. A., \& Dinis, L. 2008, Astron. Nachr., 329, 801
\bibitem[\protect\citeauthoryear{Charbonnel et al.}{1993}]{Charb93} Charbonnel, C., Meynet, G., Maeder, A., Schaller, G., \& Schaerer, D. 1993, \aaps, 101, 415
\bibitem[\protect\citeauthoryear{Clari{\'a}}{1977}]{Clar77} Clari{\'a}, J. J. 1977, \pasp, 89, 803
\bibitem[\protect\citeauthoryear{Constantine et al.}{1969}]{Const69} Constantine, S. M., Harris, B. J., \& Nikoloff, I. 1969, Proc. Astron. Soc. Aust., 1, 207
\bibitem[\protect\citeauthoryear{Deeming}{1975}]{Deem75} Deeming T. J. 1975, \apss, 36, 137
\bibitem[\protect\citeauthoryear{Dias et al.}{2001}]{Dias01} Dias, W. S., L{\'e}pine, J. R. D., \& Alessi, B. S. 2001, \aap, 376, 441
\bibitem[\protect\citeauthoryear{Dias et al.}{2002}]{Dias02} Dias, W. S., Alessi, B. S., Moitinho, A., \& L{\'e}pine, J. R. D. 2002, \aap, 389, 871
\bibitem[\protect\citeauthoryear{Dias et al.}{2006}]{Dias06} Dias, W. S., Assafin, M., Fl{\'o}rio, V., Alessi, B. S., \& L{\'i}bero, V. 2006, \aap, 446, 949
\bibitem[\protect\citeauthoryear{Francic}{1989}]{Franc89} Francic, S. P. 1989, \aj, 98, 888
\bibitem[\protect\citeauthoryear{Gieseking}{1985}]{Gies85} Gieseking, F. 1985, \aaps, 61, 75
\bibitem[\protect\citeauthoryear{Gonz{\'a}lez \& Lapasset}{2000}]{Gonz00} Gonz{\'a}lez, J. F., \& Lapasset, E. 2000, \aj, 119, 2296
\bibitem[\protect\citeauthoryear{Gonz{\'a}lez \& Lapasset}{2001}]{Gonz01} Gonz{\'a}lez, J. F., \& Lapasset, E. 2001, \aj, 121, 2657
\bibitem[\protect\citeauthoryear{Guthnik \& Prager}{1914}]{Guth14} Guthnik, P., \& Prager, R. 1914, Ver{\"o}ffentl. der k{\"o}nigl. Sternwarte Berlin-Babelsberg, 1
\bibitem[\protect\citeauthoryear{Hensberge et al.}{1977}]{Hensb77} Hensberge, H., de Loore, C., Zuiderwijk, E. J., \& Hammerschlag-Hensberge, G. 1977, \aap, 54, 443
\bibitem[\protect\citeauthoryear{Javakhishvili et al.}{2006}]{Javak06} Javakhishvili, G., Kukhianidze, V., Todua, M., \& Inasaridze, R. 2006, \aap, 447, 915
\bibitem[\protect\citeauthoryear{Kharchenko et al.}{2004}]{Khar04} Kharchenko, N. V., Piskunov, A. E., R{\"o}ser, S., Schilbach, E., \& Scholz, R.-D. 2004, Astron. Nachr. 325, 740
\bibitem[\protect\citeauthoryear{King}{1978a}]{King78a} King, D. S., Journal and Proceedings of The Royal Society of New South Wales, 111, 1
\bibitem[\protect\citeauthoryear{King}{1978b}]{King78b} King, D. S., Journal and Proceedings of The Royal Society of New South Wales, 111, 61
\bibitem[\protect\citeauthoryear{King}{1980}]{King80} King, D. S., Journal and Proceedings of The Royal Society of New South Wales, 113, 7
\bibitem[\protect\citeauthoryear{Koen \& Eyer}{2002}]{Koen02}Koen, C., \& Eyer, L. 2002, \mnras, 331, 45
\bibitem[\protect\citeauthoryear{Lafler \& Kinman}{1981}]{Lafl76} Lafler, J., \& Kinman, T. D. 1976, \apjs, 11, 216
\bibitem[\protect\citeauthoryear{Landstreet et al.}{2007}]{Land07} Landstreet, J. D., Bagnulo, S., Andretta, V., et al. 2007, \aap, 470, 685
\bibitem[\protect\citeauthoryear{Landstreet et al.}{2008}]{Land08} Landstreet, J. D., Silaj, J., Andretta, V., et al. 2008, \aap, 481, 465
\bibitem[\protect\citeauthoryear{Lehmann et al.}{2007}]{Lehm07} Lehmann, H., Tkachenko, A., Fraga, L., Tsymbal, V., \& Mkrtichian, D. E. 2007, \aap, 471, 941
\bibitem[\protect\citeauthoryear{Luck et al.}{2000}]{Luck00} Luck, R. E., Andrievsky, S. M., Kovtyukh, V. V., Korotin, S. A., \& Beletsky, Yu. V. 2000, \aap, 361, 189
\bibitem[\protect\citeauthoryear{Maitzen}{1976}]{Mai76} Maitzen, H. M. 1976, \aap, 51, 223
\bibitem[\protect\citeauthoryear{Maitzen}{1980}]{Mai80} Maitzen, H. M. 1980, \aap, 89, 230
\bibitem[\protect\citeauthoryear{Maitzen}{1993}]{Mai93} Maitzen, H. M. 1993, \aaps, 102, 1
\bibitem[\protect\citeauthoryear{Maitzen \& Hensberge}{1981}]{MH81} Maitzen, H. M., \& Hensberge, H. 1981, \aap, 96, 151
\bibitem[\protect\citeauthoryear{Maitzen et al.}{1980}]{MWW80} Maitzen, H. M., Weiss, W. W., \& Wood, H. J. 1980, \aap, 81, 323
\bibitem[\protect\citeauthoryear{Manfroid \& Mathys}{2000}]{Manf00} Manfroid, J., \& Mathys, G. 2000, \aap, 364, 689
\bibitem[\protect\citeauthoryear{Mathys et al.}{1989}]{Mathy89} Mathys, G., Maitzen, H. M., North, P., et al. 1989, The Messenger, 55, 41
\bibitem[\protect\citeauthoryear{Mathys et al.}{1997}]{Mathy97} Mathys, G., Hubrig, S., Landstreet, J. D., Lanz, T., Manfroid, J. 1997, \aaps, 123, 353
\bibitem[\protect\citeauthoryear{Mermilliod et al.}{2008}]{Mer08} Mermilliod, J.-C., Mayor, M., \& Udry, S. 2008, \aap, 485, 303
\bibitem[\protect\citeauthoryear{Mikul{\'a}\v{s}ek et al.}{2007}]{Miku07} Mikul{\'a}\v{s}ek, Z., Janik, J., Zverko, 
J., et al. 2007, Astron. Nachr., 328, 10
\bibitem[\protect\citeauthoryear{Mikul{\'a}\v{s}ek et al.}{2010}]{Miku10} Mikul{\'a}\v{s}ek, Z., Krti\v{c}ka, J., Henry, G. W., et al. 2010, \aap,
511, L7
\bibitem[\protect\citeauthoryear{Netopil et al.}{2008}]{Net08} Netopil, M., Paunzen, E., Maitzen, H. M., North, P., \& Hubrig, S. 2008, \aap, 491, 545
\bibitem[\protect\citeauthoryear{North}{1987}]{North87} North, P. 1987, \aaps, 70, 247
\bibitem[\protect\citeauthoryear{North et al.}{1988}]{North88} North, P., Jasniewicz, G., \& Waelkens, Ch. 1988, IBVS, 3199
\bibitem[\protect\citeauthoryear{North et al.}{1992}]{North92} North, P., Brown, D. N., \& Landstreet, J. D. 1992, \aap, 258, 389
\bibitem[\protect\citeauthoryear{P{\"o}hnl et al.}{2005}]{Poehn05} P{\"o}hnl, H., Paunzen, E., \& Maitzen, H. M. 2005, \aap, 441, 1111
\bibitem[\protect\citeauthoryear{Paunzen \& Netopil}{2006}]{PN06} Paunzen, E., \& Netopil, M. 2006, \mnras, 371, 1641
\bibitem[\protect\citeauthoryear{Renson}{1978}]{Rens78} Renson, P. 1978, \aap, 63, 125 
\bibitem[\protect\citeauthoryear{Renson \& Manfroid}{2009}]{Rens09} Renson, P., \& Manfroid, J. 2009, \aap, 498, 961
\bibitem[\protect\citeauthoryear{Schaerer et al.}{1993a}]{Schaer93a} Schaerer, D., Meynet, G., Maeder, A., \& Schaller, G. 1993a, \aaps, 98, 523
\bibitem[\protect\citeauthoryear{Schaerer et al.}{1993b}]{Schaer93b} Schaerer, D., Charbonnel, C., Meynet G., Maeder, A., \& Schaller, G. 1993b, \aaps, 102, 339   
\bibitem[\protect\citeauthoryear{Schaller et al.}{1992}]{Schall92} Schaller, G., Schaerer, G., Meynet, G., \& Maeder, A. 1992, \aaps, 96, 269 
\bibitem[\protect\citeauthoryear{Stellingwerf}{1978}]{Stell78} Stellingwerf, R. F. 1978, \apj, 224, 953
\bibitem[\protect\citeauthoryear{Stibbs}{1950}]{Stibbs50} Stibbs, D. W. N. 1950, \mnras, 110, 395
\bibitem[\protect\citeauthoryear{van Leeuwen et al.}{1997}]{Leeuw97} van Leeuwen, F., Evans, D. W., Grenon, M., et al. 1997, \aap, 323, L61
\bibitem[\protect\citeauthoryear{van Leeuwen}{2007}]{Leeuw07} Van Leeuwen, F. 2007, \aap, 474, 653
\bibitem[\protect\citeauthoryear{Warhurst}{2004}]{War04} Warhurst, P. M. 2004, Baltic Astronomy, 13, 597
\bibitem[\protect\citeauthoryear{Zacharias et al.}{2009}]{Zach09} Zacharias, N. et al. 2009, UCAC3 Catalogue (VizieR On-line Data Catalog I/315)
\end{thebibliography}
\end{document}